\documentstyle[aps,graphicx,tighten,preprint]{revtex}
\begin{document}
\title{Quantum cryptography with a predetermined key, using continuous variable 
Einstein-Podolsky-Rosen correlations\\ }
 \vskip 1 truecm
\author{M. D. Reid\\ }

\address{Physics Department, University of Queensland, Brisbane, Australia\\ }  	  
\date{\today}
\maketitle
\vskip 1 truecm
\begin{abstract}
Correlations of the type discussed by EPR in their original 1935 
paradox for continuous variables exist for the quadrature 
phase amplitudes of two spatially separated 
fields. These correlations were experimentally reported in 1992. We 
propose to use such EPR beams in quantum cryptography, 
to transmit with high efficiency messages in such a way that the 
receiver and sender may later determine whether eavesdropping has 
occurred. The merit of the new proposal is in the possibility of 
transmitting a reasonably secure yet predetermined key. This would allow relay of 
a cryptographic key over long distances in the presence of lossy 
channels.
\end{abstract}
\narrowtext
\vskip 0.5 truecm
Intriguing is the possibility of using quantum mechanics to transmit 
signals in a way that any eavesdropping can be detected by the receiver 
and sender. 
This new field of quantum cryptography $^{\cite{1,2}}$ 
has attracted much attention.

In the pioneering proposal of Bennett and Brassard $^{\cite{1}}$ the sender 
(Alice) transmits to the receiver (Bob)  
photon pulses in one of two orthogonal polarisations (labeled $0$ and 
$1$), where the 
orientation (basis) of polarisation randomly shifts between 
$0^{o}$ and $45^{o}$. 
The $0,1$ choice of polarisation represents the bit value. 
Bob randomly selects a basis ($0^{o}$ or $45^{o}$) for a 
polarisation measurement, and records the resulting bit value.  
Alice and Bob  
later compare notes, through a public channel, on the sequence of 
orientations ($0^{o}$ or $45^{o}$) 
 chosen. The bit sequence where Bob 
selected the same orientation as Alice forms a key, to be used later  
to encrypt messages. 
While classically an eavesdropper could measure 
with perfect accuracy components of polarisation along both 
directions, quantum mechanics 
forbids this by way of the uncertainty principle. 
 As a consequence the eavesdropper cannot always 
regenerate the original state transmitted by Alice. 
The resulting discrepancy between the results recorded by Alice and 
Bob gives warning to the interference by the eavesdropper. No 
discrepancy implies a secure key.

Other proposals $^{\cite{2}}$, such as that suggested by Ekert, propose
 to use a sequence of two spatially 
separated photons with correlated polarisation, and 
whose joint polarisation measurements are predicted by 
quantum mechanics to show a violation of a Bell  
inequality $^{\cite{3}}$. Such fields have no local hidden variable 
interpretation. Any measurement, and subsequent state regeneration to 
mask interference, by an eavesdropper along one of these 
two channels will alter the statistics 
so that a Bell inequality is always satisfied.
Again a fundamental aspect of quantum mechanics is utilized to alert 
receiver and sender to eavesdropping.

The majority of proposals so far focus on the use of single photons to transmit 
information. 
A significant current limitation to the practicality of such schemes is the 
poor efficiency of photon counting detectors. This contributes to a 
significant loss factor which makes direct efficient communication of 
sequences predetermined by Alice  
difficult. Photon-based proposals rely in practice on establishing
 a sequence (key)   a posteriori 
 from infrequent detected photons. 

 Recently Ralph $^{\cite{4}}$  and Hillery $^{\cite{5}}$  have suggested 
 cryptographic schemes based on measurement of (continuous 
variable) field 
quadrature phase amplitudes. In their proposals Alice transmits a bit 
value by way of   
squeezed signals, which means that the fluctuation in one quadrature 
phase amplitude is reduced to a level below that corresponding to the 
standard quantum limit as determined by the uncertainty principle. 
Security is provided as a result of the uncertainty principle since an 
eavesdropper (Eve) cannot measure both noncommuting quadrature 
amplitudes to arbitrary  
accuracy. As a result Bob's signal after Eve's 
interference will contain extra noise, detectable when Alice and Bob 
compare the bit values received by Bob with the bit values sent by 
Alice. In this way, following the example of Bennett and Brassard, a 
secure key can be established.

In this paper it is suggested to use continuous variable measurements 
in such a way so as to allow transmission of a predetermined sequence  
(or key) directly  from sender to receiver. Later, communication 
through a public channel can check whether eavesdropping has occurred. 
Security is provided not by comparison of Bob's received with Alice's sent bit 
values, but by establishing whether Einstein-Podolsky-Rosen 
correlations $^{\cite{6}}$ between two beams, one retained by Alice and the other 
transmitted with signal to Bob, are maintained after transmission. 
In this last respect the proposal is not unlike the photon-based proposal
 of Ekert where 
security is based on the confirmation by Alice and Bob of a violation of a Bell 
inequality. 
 
The scheme involves only quadrature phase amplitude measurements, 
which can be performed with high efficiency. The predetermined nature 
of the sequence takes most advantage of this high efficiency, since 
every bit value sent can contribute to the final message. This 
contrasts with previous 
schemes for which part of the sequence, randomly selected after 
transmission, is used only 
to establish security by way of the public channel.

 The predetermined nature 
of the sequence could also aid incorporation of special repeaters, 
where the signal and correlated beams are regenerated to 
help compensate for transmission loss. 
This method could potentially secure a single key between a single 
sender-receiver pair a long distance apart.

  Correlations of the type discussed by Einstein, Podolsky and Rosen 
  (EPR) in their original 1935 
paradox $^{\cite{6}}$, for continuous variables, exist for the quadrature 
phase amplitudes of two spatially separated 
fields $^{\cite{7}}$.
The technology of quadrature phase amplitude measurement is 
sufficiently advanced that in 1992 these correlations were detected, 
without detection 
efficiency problems, by Ou et al $^{\cite{8}}$.
Such EPR correlated beams have 
recently been utilized to enable 
quantum state teleportation with continuous variables $^{\cite{9}}$. 
Further work $^{\cite{10}}$ has shown  
that quadrature phase amplitude 
measurements on certain twin beams can predict violations of Bell 
inequalities. 

 Consider the nondegenerate parametric down conversion process, 
 modeled by two field modes with boson 
 operators $\hat{a}$ and $\hat{b}$, with the interaction 
 Hamiltonian 
 $H_I = i\hbar \kappa(\hat{a}^{\dagger} \hat{b}^{\dagger} -\hat{a}\hat{b})$. 
 We define the quadrature phase amplitudes $
		\hat{X}_a =(\hat{a}+\hat{a}^\dagger)$, $
		\hat{P}_a= (\hat{a}-\hat{a}^\dagger)/i$, 
		$\hat{X}_b =(\hat{b}+\hat{b}^\dagger)$ and $
		\hat{P}_b= (\hat{b}-\hat{b}^\dagger)/i$.
The Heisenberg 
uncertainty relation for the orthogonal 
amplitudes of mode $\hat{a}$ is 
 $\Delta^{2}X_{a}¥\Delta^{2}P_{a}¥\geq  1$.  
 The output quadrature amplitudes are  
 \begin{eqnarray}
 	\hat{X}_a (t) & = & \hat{X}_a (0)cosh(\kappa t) + 
	\hat{X}_b (0)sinh(\kappa t) \nonumber \\
	\hat{X}_b (t) & = & \hat{X}_b (0)cosh(\kappa t) + 
	\hat{X}_a (0)sinh(\kappa t) \nonumber \\
	\hat{P}_a (t) & = & \hat{P}_a (0)cosh(\kappa t) - 
	\hat{P}_b (0)sinh(\kappa t) \nonumber \\
	\hat{P}_b (t) & = & \hat{P}_b (0)cosh(\kappa t) - 
	\hat{P}_a (0)sinh(\kappa t).
\label{eqn:X(T)}
\end{eqnarray}
where $\kappa$ is proportional to the strength of parametric 
interaction 
and the $t=0$ operators represent inputs. 
As $\kappa t$ increases, 
$\hat{X}_a(t)$ becomes increasingly correlated with $\hat{X}_b(t)$, and 
$\hat{P}_a(t)$ becomes increasingly correlated with $-\hat{P}_b(t)$, the 
correlation becoming perfect in the limit $\kappa 
T\rightarrow\infty$. With output fields $\hat{a}$ and $\hat{b}$ 
spatially separated, this is the situation $^{\cite{7}}$ of the 1935 
EPR correlations. 

For imperfect correlation, the degree of 
correlation may still be sufficient to ensure EPR correlations 
$^{\cite{7}}$. 
The results for measurements $\hat{X}_a(t)$ and $\hat{X}_b(t)$ (or 
$\hat{P}_a(t)$ and $\hat{P}_b(t)$) can be compared, yielding an 
estimate of the error in inferring the result of 
measurement $\hat{X}_a(t)$ 
on mode $\hat{a}$, based on a 
measurement $\hat{X}_b(t)$ on mode $\hat{b}$. We calculate $
 	\delta_{x} =  \hat{X}_a(t)-\gamma\hat{X}_b(t)$ and $
 	\delta_{p} =  \hat{P}_a(t)+\gamma\hat{P}_b(t)$, where the factor $\gamma$ may be  
modified to give the minimum error. 
One can calculate the variances associated with the inference 
of $\hat{X}_{a}$ from $\gamma\hat{X}_{b}$, and $\hat{P}_{a}$ from 
$\gamma\hat{P}_{b}$:
$\Delta_{x,inf}^{2}  =  <{\delta_{x}^{2}}>-<{\delta_{x}}>^{2}$ and $
	\Delta_{p,inf}^{2}  =  <{\delta_{p}^{2}}>-<{\delta_{p}}>^{2}$. 
The minimum variance $\Delta_{x,inf,min}^{2}$ (and 
 $\Delta_{p, inf,\mbox{min}}^{2})$ occurs for a particular value of $\gamma$. 
Finding the turning point with $\gamma$ yields 
(with $\gamma=<{\hat{X}_a(T)},{\hat{X}_b(T)}>/\Delta^{2}¥{\hat{X}_b(T)}$) 
$\Delta_{x,inf,min}^{2}= 
	\frac{\Delta{\hat{X}^{2}_a(T)}\Delta{\hat{X}^{2}_b(T)}
	-[<{\hat{X}_a(T)},{\hat{X}_b(T)}>]^{2}}
	{\Delta{\hat{X}^{2}_b(T)}}$, where $<x,y>= <xy>-<x><y>$ and one deduces a $\Delta_{p,inf,min}^{2}$ 
in similar fashion. 
  
EPR correlations are obtained when the product  
$\Delta_{x,inf}^{2}\Delta_{p,inf}^{2}$ drops below the 
quantum limit given by $\Delta^{2}X_{a}¥\Delta^{2}P_{a}¥\geq  1$ 
$^{\cite{7}}$: 
\begin{eqnarray}
\Delta_{x,inf}^{2}  \Delta_{p,inf}^{2} < 1.
	\end{eqnarray}
For arbitrary coherent input states, we predict from 
(1) $^{\cite{7}}$ ($\gamma=\tanh 2\kappa t$)
\begin{eqnarray}
	\Delta_{x,inf,min}^{2}=\Delta_{p,inf,min}^{2}=1/\cosh 
	2\kappa t
\end{eqnarray}
An identical argument and results hold if the 
measured operators are $X_{a}-<X_{a}>$, $X_{b}-<X_{b}>$, $P_{a}-<P_{a}>$ 
and $P_{b}-<P_{b}>$, the fluctuations about 
the mean, as opposed to $X_{a}$, $X_{b}$, $P_{a}$ and $P_{b}$.   
 
 With vacuum inputs to $\hat{a}$ and 
$\hat{b}$, Bob and Alice can secure a random key, using the 
potentially perfect   
correlation between quadrature amplitudes. We propose a different 
scheme, to allow for predetermined sequences, and imperfect 
correlation. For the purposes of cryptography (Figure 1), Alice chooses as  
 input to the nondegenerate parametric amplifier  
one of two possible states: 
 the input for $\hat{a}$ is either a coherent state 
 $|\alpha_{0}\exp^{i\pi/4}>_{a}$ 
 (bit value $1$) or 
 a coherent state $|\alpha_{1}\exp^{i\pi/4}¥>_{a}$ (bit value $0$), 
 where $\alpha_{0}$ and $\alpha_{1}$ are real. 
 The input for $\hat{b}$ is a vacuum state $|0>_{b}$. 
 The signal is 
 transmitted by spatially separating the two output fields and 
 propagating to Bob the output field of mode $\hat{a}$. Bob can read the 
 message by measuring either $\hat{X}_a (t)$ or $\hat{P}_a 
 (t)$. Suppose Bob chooses to measure $\hat{X}_a (t)$.
 The probability distribution for his obtaining a result $x$, given 
 Alice's choice $|\alpha_{0}\exp^{i\pi/4}>$, 
 is the gaussian 
 $\exp{[-(x-\sqrt{2}\alpha_{0}\cosh{\kappa t})^{2}/2\sigma^{2}]}/
 \sigma\sqrt{2\pi}$ with mean  $\sqrt{2}\alpha_{0}\cosh{\kappa t}$ and standard 
 deviation  
 $\sigma=\sqrt{\cosh{2\kappa t}}$.
 If Alice chose $|\alpha_{1}\exp^{i\pi/4}>$ the probability for Bob's outcome is   
 $\exp{[-(x-\sqrt{2}\alpha_{1}\cosh{\kappa t})^{2}/2\sigma^{2}]}/\sigma\sqrt{2\pi}$, 
 the gaussian mean shifted by 
  $\sqrt{2}(\alpha_{0}-\alpha_{1})\cosh{\kappa t}$. Provided 
 $\sigma\ll \sqrt{2}(\alpha_{0}-\alpha_{1})¥\cosh{\kappa t}$, the bit value 
 is clearly 
 determined from Bob's result $x$ (Figure 2): $x$ near 
 $\sqrt{2}\alpha_{0}\cosh{\kappa 
 t}$ implies $1$; $x$ near  
 $\sqrt{2}\alpha_{1}¥\cosh{\kappa t}$ implies zero. The bit value can also be 
 determined by a measurement of quadrature phase amplitude $\hat{P}_a 
 (t)$, in this case the input $|\alpha_{0}\exp^{i\pi/4}>$ giving a 
 gaussian distribution about $\sqrt{2}\alpha_{0}\cosh{\kappa 
 t}$ (bit value $1$), 
 while $|\alpha_{1}\exp^{i\pi/4}>$ gives a distribution centered 
 about $\sqrt{2}\alpha_{1}\cosh \kappa t$ (bit value $0$).   

Bob records the results of his consecutive quadrature phase measurements, 
randomly selecting to measure either $\hat{X}_a (t)$ or $\hat{P}_a (t)$, and 
subtracting from his result either $\sqrt{2}\alpha_{0}\cosh \kappa t$ or 
$\sqrt{2}\alpha_{1}\cosh \kappa t$,
 so that only the fluctuation about the mean of the  
particular distribution is recorded (Figure 2). Bob then 
communicates to Alice, through a public channel, the sequence of 
recorded fluctuations together with 
measurements ($\hat{X}_a (t)$ or $\hat{P}_a (t)$) chosen (the bit value 
itself is not communicated). Alice 
also makes a sequence of consecutive measurements  
$\hat{X}_b (t)$ or $\hat{P}_b (t)$, (preferably) 
to coincide with Bob's measurement sequence, and records similarly only the 
fluctuation about the mean (in this case 
$\sqrt{2}\alpha_{0}\sinh\kappa t$ or $\sqrt{2}\alpha_{1}\sinh\kappa t$ 
for $X_{b}$, and  
$-\sqrt{2}\alpha_{0}\sinh\kappa t$ or $-\sqrt{2}\alpha_{1}\sinh\kappa t$ 
 for $P_{b}$). Bob and Alice 
compare notes, through the public channel, to calculate a  
$\Delta_{x,inf}^{2}\Delta_{p,inf}^{2}$.    
The predicted minimum is, for 
optimized $\gamma$, given by (3).

Verification by Bob and Alice of the EPR correlations 
$\Delta_{x,inf}^{2}\Delta_{p,inf}^{2} < 1$ gives an indication  
of interference by an eavesdropper (Eve). Let us consider various practical 
options by Eve.  To determine the signal Eve's first obvious choice 
may be to capture the field $\hat a$ and measure either $\hat{X}_{a}$ or $\hat{P}_{a}$. 
If she is able to predetermine correctly for each bit value the 
choice ($\hat{X}_{a}$ or $\hat{P}_{a}$) to be made by Bob, 
Eve can make the same choice and conceal her 
eavesdropping. However Bob's choice is delayed until after his 
detection of $\hat a$ forcing errors in Eve's selection. Quantum mechanics 
makes it impossible for Eve to measure both amplitudes ($\hat{X}_{a}$ 
and $\hat{P}_{a}$) to an uncertainty better 
than that given by the Heisenberg uncertainty relation. More 
importantly, Eve 
cannot regenerate and transmit to Bob a single mode 
state with 
both well defined $\hat{X}$ and $\hat{P}$, but is limited by 
$\Delta^{2}\hat{X}\Delta^{2}\hat{P} \geq  1$. For example Eve 
may select to measure $\hat{X}_{a}$ rather precisely so that the 
error in the measurement is of order $\Delta_{m}¥^{2}=1/r$, where $r>1$.    
Eve may then generate, to transmit to Bob, 
a ``squeezed'' state with this reduced 
fluctuation in $X$, so that the new operator describing the quadrature 
measurement now made by Bob is $\hat{X}_{a}^{new}={x}_{a}+\delta 
\hat{X}_{a}$ 
where $x_{a}$ is the result of Eve's measurement and 
$\Delta^{2}\delta \hat{X}_{a}=1/r$.  Quantum mechanics compels 
an enhanced fluctuation in $\hat{P}$, so that the operator describing the 
quadrature measurement $\hat{P}_{a}¥$ made by Bob on this 
retransmitted state is 
$\hat{P}_{a}^{new}={p}_{a}+\delta \hat{P}_{a}$ where at best $\Delta^{2} 
\delta \hat{P}=r$ for a minimum 
uncertainty squeezed state. The variances $\Delta_{x,inf,min}^{2}$ and 
$\Delta_{p,inf,min}^{2}$ 
testing for supposed EPR correlations 
are now  
$\Delta_{x^{new}¥,inf,min}^{2}=\Delta_{x,inf,min}^{2}+\Delta^{2}¥ \delta 
\hat{X}_{a}¥$ 
and $\Delta_{p^{new}¥,inf,min}^{2}=\Delta_{p,inf,min}^{2}+\Delta^{2}¥ 
\delta \hat{P}$, where here we have $\Delta_{x,inf,min}^{2}= 
\Delta_{p,inf,min}^{2}=1/\cosh \kappa t$.  This gives 
$\Delta_{x,inf,min}^{2}\Delta_{p,inf,min}^{2} \geq 1$, and EPR correlations 
are lost, making a sensitive test 
for interference on $\hat{a}$. We note that it is possible for Eve to gain 
access to bit values, but whether this has occurred is later checked by 
communication between sender and receiver.

To improve her chances, as discussed by Ralph $^{\cite{4}}$, 
Eve may alternatively opt to 
make a partial interference of beam $a$ 
by tapping off only part of the beam using a partially-transmitting 
beam splitter, with $a$ and $a_{vac}$ as inputs, 
where $a_{vac}$ is a vacuum input (Figure 3). The outputs 
are:  $\hat{a}_{Bob}=\sqrt{\eta}\hat{a}+\sqrt{1-\eta}\hat{a}_{vac}$, the field 
transmitted and detected by Bob; and 
$\hat{a}_{Eve}=\sqrt{1-\eta}\hat{a}-\sqrt{\eta}\hat{a}_{vac}$, the field 
detected by Eve to allow her measurement of $X_{a}$. Here $\eta$ gives 
the fraction of photons transmitted, on to Bob, by the beamsplitter. 
We define the quadrature 
amplitudes $\hat{X}_{a}^{Bob}=\hat a_{Bob}+\hat{a}_{Bob}^\dagger$,
		$\hat{P}_{a}^{Bob}= (\hat{a}_{Bob}¥-\hat{a}_{Bob}^\dagger)/i$,  
 $\hat X_{a}^{Eve}=\hat a_{Eve}+\hat{a}_{Eve}^\dagger$ and 
 		$\hat{P}_{a}^{Eve}= (\hat{a}_{Eve}¥-\hat{a}_{Eve}^\dagger)/i$. For a 
 		vacuum input we have 
$\Delta^{2}\hat X_{vac}=\Delta^{2}\hat P_{vac}= 1$.  
  \begin {eqnarray}
\hat{X}_{a}^{Bob}(t)&=&\sqrt{\eta}X_{a}(t) + \sqrt{1-\eta}X_{vac}\nonumber\\
  \hat X_{a}^{Eve}(t)&=&\sqrt{\eta}X_{vac} - \sqrt{1-\eta}X_{a}(t)\nonumber\\
   \hat{P}_{a}^{Bob}(t)&=&\sqrt{\eta}P_{a}(t)+\sqrt{1-\eta}P_{vac}\nonumber\\
\hat{P}_{a}^{Eve}(t)&=&\sqrt{\eta}P_{vac} - \sqrt{1-\eta}P_{a}(t)
 \end{eqnarray}
 The variances $\Delta_{x,inf,min}^{2}$ and 
$\Delta_{p,inf,min}^{2}$ later measured by Alice and Bob, 
testing for EPR correlations, 
are now
\begin{eqnarray}  
\Delta_{x^{new}¥,inf,min}^{2}&=&\eta\Delta_{x,inf,min}^{2}+
(1-\eta)\Delta^{2}\hat{X}_{vac}\nonumber\\ 
\Delta_{p^{new}¥,inf,min}^{2}&=&\eta\Delta_{p,inf,min}^{2}+
(1-\eta)\Delta^{2}\hat P_{vac}
\end{eqnarray}
 With $\eta\rightarrow 1$ the back-action noise 
 ($\sqrt{1-\eta}X_{vac}$ for 
 measurement $X$) feeding into Bob's 
 signal as a result of Eve's tapping is decreased. In this limit, the change 
 $(1-\eta)\Delta^{2}\hat{X}_{vac}$ and $(1-\eta)\Delta^{2}\hat P_{vac}$ 
 to the variances $\Delta_{x,inf}^{2}$ and $\Delta_{p,inf}^{2}$ 
 respectively, as a result of Eve's eavesdropping becomes increasingly 
 undetectable. Eve however pays the price, since she observes a 
 reduced signal ($-\sqrt{1-\eta}X_{a}(t)$ for the measurement $X$) with 
 increased noise (due to $\sqrt{\eta}X_{vac}$), 
 limiting her ability to obtain 
 information from the channel. Witn noise 
 $\sqrt{\eta}X_{vac}$ from the vacuum input 
 increasing as $\eta\rightarrow 1$, a point is reached where she can 
 no longer resolve the two peaks, separated by 
 $\sqrt{2}\sqrt{1-\eta}\cosh{\kappa t}(\alpha_{0}-\alpha_{1})$, 
 giving the bit value.
    
In an effort to reduce  the feedback noise $(1-\eta)\Delta^{2}\hat{X}_{vac}$
 in Bob's signal, and to allow better resolution of the bit value for 
 larger $\eta$, 
 Eve may choose to perform a quantum 
nondemolition measurement of quadrature amplitude $\hat{X}_{a}$ 
(Figure 3). Such measurements allow accurate determination of $\hat{X}_{a}$ 
(to $\Delta^{2}\hat{X} \leq  1$) and have been achieved experimentally
 $^{\cite{11}}$. The quantum nondemolition measurement may be performed 
using the beam splitter as above (Figure 3) but 
where $a_{vac}$ is a squeezed vacuum input so that 
$\Delta^{2}\hat{X}_{vac} <  1$ (suppose $\Delta 
\hat{X}_{vac}=1/r$). 
Increased squeezing of the fluctuation in $X_{vac}¥$ 
($\Delta^{2}\hat{X}_{vac} \rightarrow 0$)  implies that 
$X_{a}^{Bob}(t)=\sqrt{\eta}X_{a}(t)$ and 
  $X_{a}^{Eve}= -\sqrt{1-\eta} X_{a}(t)$  and perfect 
  inference of $X_{a}(t)$ is obtainable 
  by Eve, without any feedback vacuum noise in the value $X_{Bob}(t)$ later measured by 
  Bob. However large fluctuations in $P_{vac}¥$ (we must have $\Delta 
  \hat{P}_{vac}=r$ to satisfy the uncertainty principle for the 
  squeezed vacuum input state) necessarily 
  create a large noise 
 in $P_{a}^{Bob}$.
 \begin{equation} 
 P_{a}^{Bob}(t)=\sqrt{\eta}P_{a}(t)+\sqrt{1-\eta} P_{vac}
 \end{equation}
 This excess noise, detectable when Bob selects to measure $P$ rather 
 than $X$, causes an increase in 
$\Delta_{p^{new}¥,inf,min}^{2}=\eta\Delta_{p,inf,min}^{2}+
(1-\eta)\Delta^{2}\hat P_{vac}$, alerting Bob to Eve's interference.

 The presence of loss due to transmission will also reduce the  
 EPR correlation. 
Loss (and detection inefficiencies) may be modeled by 
a beam splitter which mixes our signal mode  
$\hat{a}$ with a vacuum field  $\hat{a}_{vac}$ to give a new output
 at Bob's detector: $ 
	\hat{a}^{new}=\sqrt{\eta}\hat{a}+\sqrt{1-\eta}\hat{a}_{vac}.
$
Here $\eta$ is the overall efficiency factor ($\eta\rightarrow1$ for 
no loss).
The new noise levels 
measured by Bob are
\begin{eqnarray}
 \Delta^{2}_{x^{new}¥,inf,min}&=&\eta 
 ^{2}\Delta^{2}_{x,inf,min}¥+(1-\eta ^{2}¥) \nonumber\\ 
 \Delta^{2}_{p^{new}¥,inf,min}&=&\eta 
 ^{2}¥\Delta^{2}_{p,inf,min}¥+(1-\eta ^{2}¥).
 \end{eqnarray} 
 With $\eta>0$, a partial loss, EPR correlations 
 are still maintained, though decreased. For
  complete loss we obtain 
 $\Delta^{2}_{x^{new}¥,inf,min}=\Delta^{2}_{p^{new}¥,inf,min}=1$.   
 
In practice, the degree of EPR correlation for a given 
transmission line and distance would be accurately established. 
 This degree of correlation is independent 
 of Alice's bit value.
Any increase of our EPR noise indicator above this pre-evaluated level 
alerts Bob to the additional loss caused by a partial tapping of 
the channel by Eve. 

Security is also provided by comparing individual results of 
measurements made by Alice and Bob. For a given transmission line and 
loss along this line, and 
for a given bit value (based on the choice $\alpha$) the mean and 
shape (the shape is predicted to be independent of the bit value) of the 
measured distribution can also be accurately recorded. A
 specified result for the 
measurement (or fluctuation about the mean) $X_{b}$ made by Alice will 
imply a conditional probability distribution for the 
measurement (or fluctuation about 
mean) $X_{a}$ made by Bob. In the absence of loss the variance of 
this conditional distribution is $\Delta^{2}_{x,inf,min}¥$. Loss 
increases the variance by the amount given above in (7). Significant 
deviation of a result for Bob from this distribution is indication of 
Eve's presence. Importantly 
loss acts to increase noise levels in $X$ and $P$ equally. Marked 
increase, for some of the bit values sent, in the deviation of 
Bob's measurement from Alice's predicted 
result for Bob would alert Alice and Bob to the possibility of Eve 
having performed a quantum nondemolition measurement as discussed above. 

Eve's best chance then may be to perform measurement with a partial beam 
splitter with standard vacuum input, in the hope that the extra noise put back 
into Bob's channel will not be noticeable over loss. To safeguard 
against this Alice and Bob must evaluate by measurements the minimum 
extra noise, or additional loss, for which they would conclude the 
existence of a potential eavesdropper. With this value of $\eta$ Eve 
 could have performed a measurement (4) and would be compelled to 
 infer a bit value based on extra noise levels as indicated by (4). 
 Bob and Alice must select the difference between inputs $\alpha_{0}$ 
 and $\alpha_{1}$ so that Eve is unable to resolve the bit value with 
 this extra noise.  

 Schemes 
using the violation of a bell inequality $^{\cite{2}}$ can 
also be proposed for continuous variable quadrature phase 
detection, since the failure of local realism has recently $^{\cite{10}}$ been 
predicted possible for such measurements, for certain types of quantum states. 
One such state is the pair-coherent state $^{\cite{10}}$   
 \begin{equation}
	|\Psi> = N \int_{0}^{2\pi} 
	|r_{0}e^{i\varsigma}>_{a}
	|r_{0}e^{-i\varsigma}>_{b} d\varsigma 
	\label{eqn:circle state}
\end{equation}
Here $N$ is a normalization coefficient, we choose $r_{0}=1.1$ and  
$|\alpha>_q$ ($q=a,b$) is a coherent state for the mode $\hat{q}$. 
Also we might consider the two-mode ``Schrodinger cat" state undergoing 
interaction for a time $t$ with a parametric amplifier $^{\cite{10}}$ 
\begin{equation}
	|\Psi> =N\hat{U}\left( |\alpha_{0}>_{a}|\beta_{0}>_{b}
		+|-\alpha_{0}>_{a}|-\beta_{0}>_{b}\right)
\end{equation}
where $U=\exp{[-i\hat{H_I} t/\hbar]}$, and we choose $\alpha_{0}=\beta_{0}=0.9$ and $\kappa t=0.6$ 
Our protocol is not a direct parallel of Ekert's for 
spin-$1/2$ particles, because for states (8) and (9) there is not a 
perfect correlation between quadrature amplitude measurements on 
$\hat{a},\hat{b}$.
 
 After generation of the state (8) (or (9)), the two fields $\hat{a}$ and $\hat{b}$ 
 are spatially separated. Alice may then choose to phase shift 
 the field $\hat{a}$ by $180^{o}$ or not, this choice of relative phase between $\hat{a}$ and $\hat{b}$ 
 being her signal. The field $\hat{a}$ is then propagated to Bob at 
 a distant location $A$. 
The signal is transmitted from 
Alice to Bob in the form of  blocks, consisting  
of many ($N$ say where $N$ is large) 
identical states with the same value of phase shift. Bob measures 
at a location $A$ a quadrature phase amplitude $\hat{X}_{\theta}^{A} 
=\hat{X}_{a}¥\cos{\theta}+\hat{P}_{a}\sin{\theta}$ 
 for each state comprising a certain block, where $\theta$ randomly 
varies between $\theta=0, \pi/2, 3\pi/2$, for state (8) (or
between $\theta=0, 0.42\pi, -0.28\pi, 1.42\pi, 0.72\pi$ for state (9)). 
Alice also makes a 
series of measurements 
$\hat{X}_{\phi}^B=\hat{X}_{b}¥\cos{\phi}+\hat{P}_{b}\sin{\phi}$ 
at a location $B$, 
where $\phi$ randomly varies between  $\phi=0,-\pi/4,-3\pi/4$, for 
state (8) (or between $\phi=0, -0.28\pi, 0.42\pi$ for state (9)). 
Alice 
then communicates to Bob through a public channel the 
results for her quadrature phase amplitude measurements.

Bob may build up, for each block, the probability 
distribution $P(q_{a},q_{b})$ for getting results $q_{a}$ and $q_{b}$ upon 
measurement of $\hat{X}_{a}¥$ at $\hat{a}$ and $\hat{X}_{b}¥$ at $\hat{b}$ respectively. This 
information is given by the $\theta=0$ and $\phi=0$ measurements. The 
shape of the distribution changes with the choice of phase shift, and 
gives the bit value. This information is not determinable 
from the measurements of amplitudes made on $\hat{b}$ alone, and hence 
cannot be determined by the information passed along the public 
channel. 

To check whether eavesdropping has occurred, Bob tests for 
a  Bell inequality.  
The result of the 
measurement is classified as $+1$ if the quadrature phase result $x$ is 
greater than or 
equal to zero, and $-1$ otherwise.  We define the 
probability distributions: $P_{+}^{A}(\theta)$ 
for obtaining  
$+1$ at $\hat{a}$ upon measurement of $\hat{X}_{\theta}^{A}$; $P_{+}^{B}(\phi)$ 
for obtaining $+1$ at $\hat{b}$ upon measurement of $\hat{X}_{\phi}^{B}$; and 
$P_{++}^{AB}(\theta,\phi)$ the joint probability of obtaining a $+1$ 
result at both $\hat{a}$ and $\hat{b}$. 
  The existence of a local hidden variable theory implies the
 ``strong'' Bell-Clauser-Horne inequality $^{\cite{3}}$.
  \begin{eqnarray}
 S={{P_{++}^{AB}(\theta,\phi)-P_{++}^{AB}(\theta,\phi')+P_{++}^{AB}(\theta',\phi)
 +P_{++}^{AB}(\theta',\phi')}\over{P_{+}^{A}(\theta')+P_{+}^{B}(\phi)}} \leq 1
	\label{eqnbell}
\end{eqnarray}
 For state (8), a violation of 
 this inequality occurs with $S \approx 1.0157$, and 
 with angles given by 
$\theta=0,\phi=-\pi/4,\theta'=\pi/2,\phi'=-3\pi/4$ $^{\cite{10}}$. 
For state (9), violation given by $S=1.008$ is obtained for angles 
$\theta=0.42\pi,\phi=-0.28\pi,\theta'=\-0.28\pi,\phi'=0.42\pi$ 
$^{\cite{10}}$. The 
above  
violations also hold for the states generated by phase shifting $\hat{a}$ 
by $180^{o}$, with the choice of angles for $\phi$ as before, 
but replacing $\theta$ with $\theta+\pi$ and  $\theta'$ with 
$\theta'+\pi$.

Violation of the Bell inequality at the level predicted by quantum 
mechanics ensures that no interference by Eve has  
occurred along $\hat{a}$ (see Ekert 
$^{\cite{2}}$).  
Suppose Eve performs a measurement on the 
field $\hat{a}$, measuring $\hat{X}_{\theta_{0}¥}^{A}$ say to obtain a result 
$x_{\theta_{0}¥}¥$. She then generates and transmits to Bob a state 
$|\Phi_{x_{\theta_{0}},\theta_{0}}¥>$. 
The density operator for the new combined system is 
	$\rho = \rho_{x_{\theta_{0}},\theta_{0}}^{B} 
	\rho_{x_{\theta_{0}},\theta_{0}}^{A}$
where 
$\rho_{x_{\theta_{0}¥},\theta_{0}}^{B}=
<x_{\theta_{0}¥}¥|\Psi><\Psi|x_{\theta_{0}¥}¥>$ is the reduced density 
matrix for field $\hat{b}$ given the measurement by Eve, 
$|x_{\theta_{0}¥}¥>$ 
is the eigenstate of $\hat{X}_{\theta_{0}¥}^{A}$, and 
$\rho_{x_{\theta_{0}},\theta_{0}¥}^{A}=
|\Phi_{x_{\theta_{0}},\theta_{0}¥}¥><\Phi_{x_{\theta_{0}¥},\theta_{0}¥}¥|$. 
Bob tests for the 
Bell inequality using $P_{x,y}^{AB}(\theta,\phi)$, the joint 
probability for respective results $x$ and $y$ for measurements 
$\hat{X}_{\theta}^{A}$ and $\hat{X}_{\phi}^{B}$. With intervention,   
  \begin{eqnarray}
P_{x,y}^{AB}(\theta,\phi)=\sum_{x_{\theta_{0}¥}¥}\sum_{\theta_{0}}P(x_{\theta_{0}¥}, \theta_{0}) \nonumber \\ 
<y_{\phi}|<x_{\theta_{0}}¥|\Psi><\Psi|x_{\theta_{0}¥}¥>  
|y_{\phi}¥>   \nonumber \\
<x_{\theta}¥|\Phi_{x_{\theta_{0}},\theta_{0}¥}¥><\Phi_{x_{\theta_{0}},\theta_{0}¥}|x_{\theta}¥>
  \end{eqnarray}
where $P(x_{\theta_{0}¥}, \theta_{0})$ is the probability that Eve 
obtains a result $x_{\theta_{0}}$ for her measurement. 
We have the form $
P_{x,y}^{AB}(\theta ,\phi )= \int \rho(\lambda) \quad p_{x}^A(\theta, \lambda ) 
p_{y}^B(\phi, \lambda )\quad d\lambda  
$ from which a Bell inequality follows, regardless of the 
state regenerated by Eve.

In terms of feasibility, the second scheme based on 
the Bell inequality is more likely 
to be limited by difficulty of state preparation and 
susceptibility to loss ($\eta=0.96$ destroys 
violations $^{\cite{10}}$)and is greatly limited by its use of 
redundancy.

The first scheme, not so limited, may 
offer advantages over schemes utilizing photon counting. 
The high detection 
efficiencies give a very much reduced overall loss factor, which may make it 
possible to transmit directly and efficiently a predetermined
message, later 
checking providing a means to check 
security. The generation and detection of EPR correlations with 
 $\Delta^{2}_{x,inf}\Delta^{2}_{p,inf}=0.7$ has been achieved 
 $^{\cite{8}}$. 
The 
generation of squeezed (where $\Delta^{2}\hat{X}_{\theta}^{A}<1$ for some 
$\theta$) optical and soliton pulses $^{\cite{12}}$ opens up possibilities for 
transmission of EPR correlated fields. 
The robustness of squeezing to propagation loss has not been 
keenly explored, but similar distances 
 should be achievable for EPR correlations. This loss represents the
  chief limitation 
 to long distance transmission, since loss acts to degrade the EPR 
correlations which must be kept at 
$\Delta^{2}_{x,inf}\Delta^{2}_{p,inf}<1$. 
Repeated detection and regeneration of the signal with new EPR fields 
could help combat loss. Security then relies on a set of 
senders and receivers being able to communicate reliably at a later 
stage, after the detections. 
 
 In recent applications $^{\cite{9}}$ EPR beams have been generated 
as the two 
outputs of a beam splitter with inputs a squeezed vacuum state.
 It would be possible 
to use 
such EPR systems for our cryptography scheme where the 
squeezed vacuum is 
replaced by an amplitude squeezed state.

\begin{figure}
\includegraphics[scale=0.9]{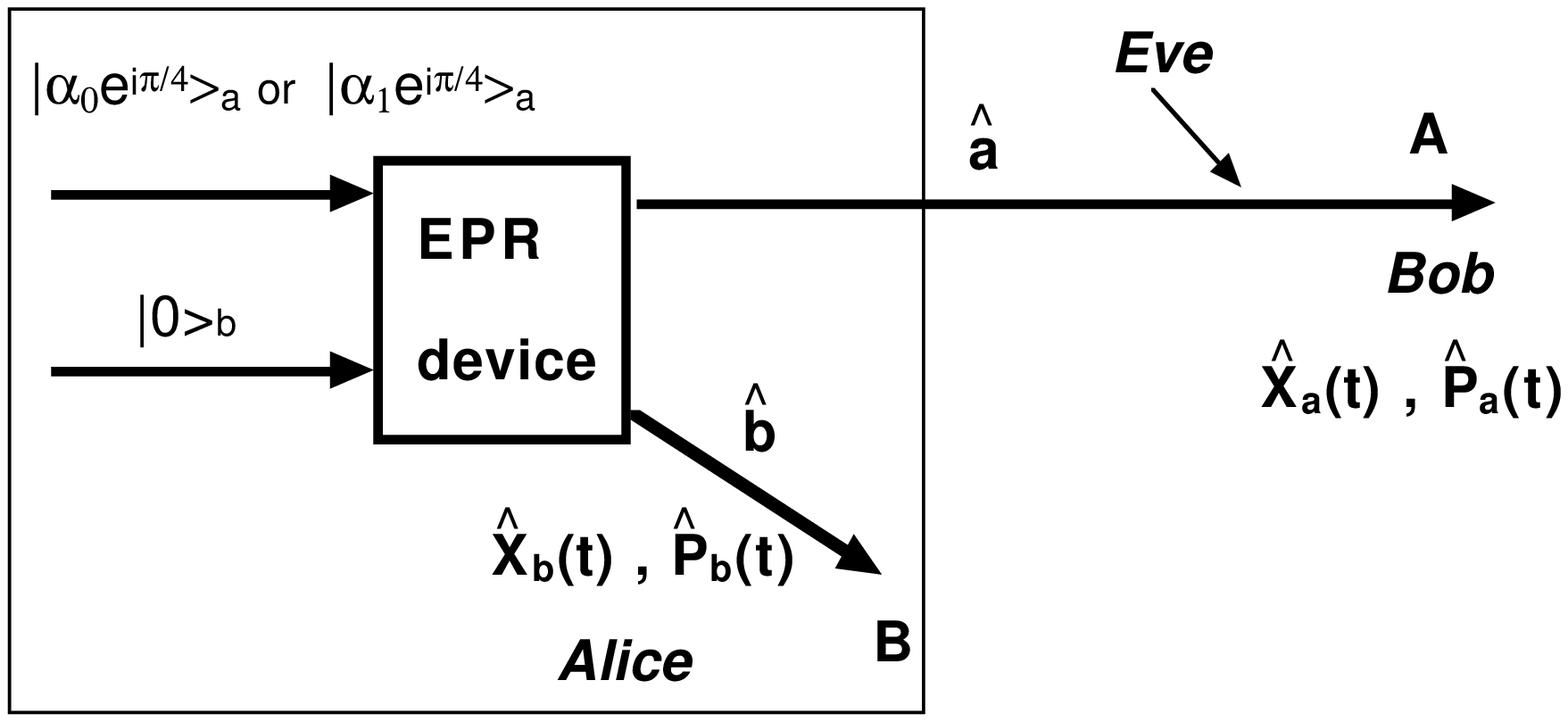}
\caption
{%
Schematic representation of the EPR cryptographic scheme. The EPR 
device generates fields $\hat{a}$ and $\hat{b}$ which are EPR correlated.
 The bit value is given by Alice's choice of 
input to $\hat{a}$.%
}
\label{fig1}
\end{figure}

\begin{figure}
\includegraphics[scale=.9]{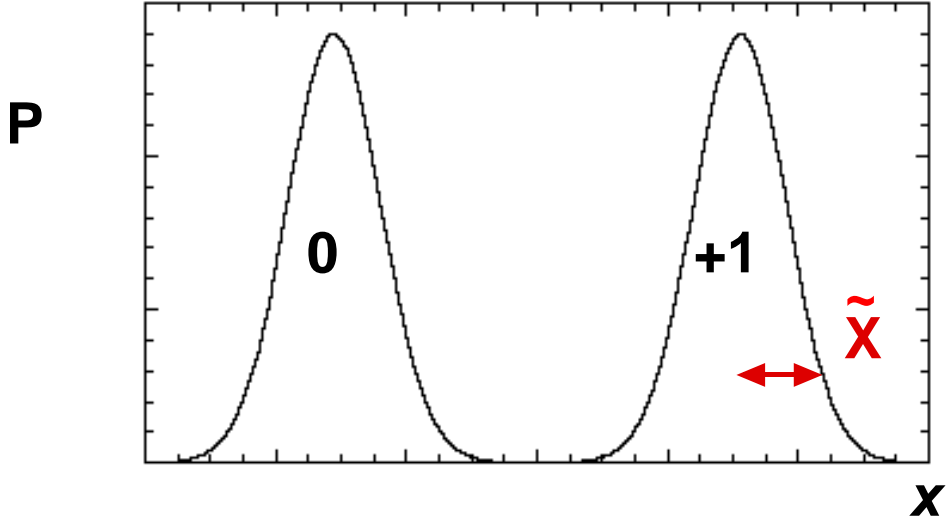}
 \caption
{%
Schematic plot of the  probability distribution $P=P(x)$ 
for obtaining a result $x$ upon measurement of the quadrature phase 
amplitude of $a$ or $b$, where one gaussian peak represents input 
 $|\alpha_{0}\exp^{i\pi/4}>_{a}$ 
 (bit value $1$)  and the other input 
 $|\alpha_{1}\exp^{i\pi/4}¥>_{a}$ (bit value $0$). 
 Bob is able to infer 
the bit value from $x$ and record, for later communication to Alice, the 
deviation $\tilde X$ of his result from 
the (known) mean of the distribution as indicated.%
}
\end{figure}

\begin{figure}
\includegraphics{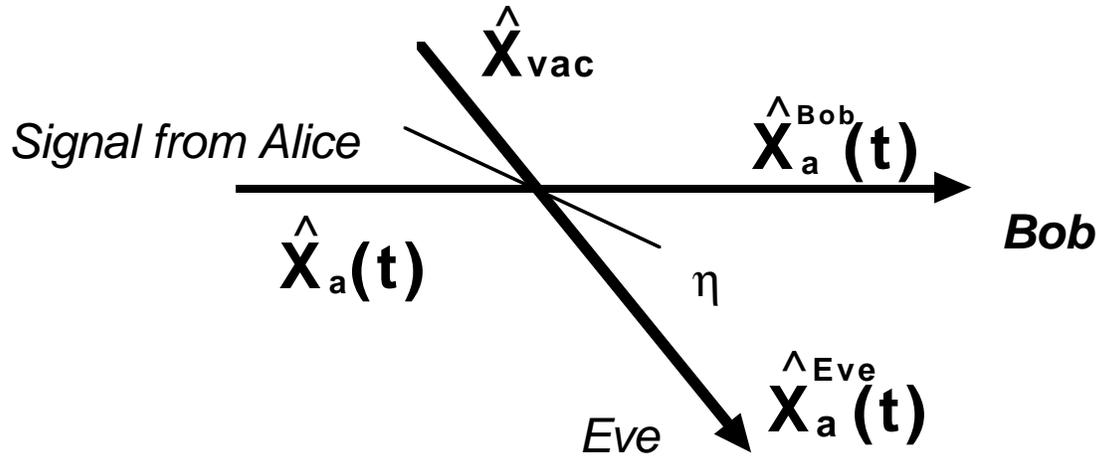}
 \caption
{%
Schematic representation of Eve's attempt to make measurement of 
$X_{a}(t)$ using a partial beam splitter.%
}
\end{figure}

\end{document}